\documentclass{aastex631}

\usepackage{amsmath}
\usepackage{amsfonts}
\usepackage{amssymb}
\usepackage{placeins}
\usepackage{hyperref}
\usepackage{multirow}

\usepackage{xcolor}
\usepackage{float}

\begin{document}

\newcommand{\pzvi}[1]{#1}
\newcommand{\revpz}[1]{#1}

\title{Implementation of a Near-Realtime Recording and Reporting System of Solar Radio Bursts}

\author[0000-0001-6855-5799]{Peijin Zhang}
\affiliation{Center for Solar-Terrestrial Research, New Jersey Institute of Technology, Newark, NJ 07102, USA}
\author[0009-0001-6649-5195]{Anastasia Kuske}
\affiliation{Center for Solar-Terrestrial Research, New Jersey Institute of Technology, Newark, NJ 07102, USA}
\author[0000-0002-0660-3350]{Bin Chen}
\affiliation{Center for Solar-Terrestrial Research, New Jersey Institute of Technology, Newark, NJ 07102, USA}
\author[0000-0002-3627-8261]{Mengjia Xu}
\affiliation{Department of Data Science, New Jersey Institute of Technology, Newark, NJ 07102, USA}
\author{Gelu Nita}
\affiliation{Center for Solar-Terrestrial Research, New Jersey Institute of Technology, Newark, NJ 07102, USA}

\author{Marin M. Anderson}
\affiliation{Owens Valley Radio Observatory, California Institute of Technology, Big Pine, CA 93513, USA}
\affiliation{Jet Propulsion Laboratory, California Institute of Technology, Pasadena, CA 91011, USA}

\author{Judd D. Bowman}
\affiliation{School of Earth and Space Exploration, Arizona State University, Tempe, AZ 85287, USA}

\author{Ruby Byrne}
\affiliation{Cahill Center for Astronomy and Astrophysics, California Institute of Technology, Pasadena, CA 91125, USA}
\affiliation{Owens Valley Radio Observatory, California Institute of Technology, Big Pine, CA 93513, USA}

\author{Morgan Catha}
\affiliation{Owens Valley Radio Observatory, California Institute of Technology, Big Pine, CA 93513, USA}

\author{Xingyao Chen}
\affiliation{Center for Solar-Terrestrial Research, New Jersey Institute of Technology, Newark, NJ 07102, USA}

\author{Sherry Chhabra}
\affiliation{Center for Solar-Terrestrial Research, New Jersey Institute of Technology, Newark, NJ 07102, USA}
\affiliation{George Mason University, Fairfax, VA 22030, USA}

\author{Larry D'Addario}
\affiliation{Cahill Center for Astronomy and Astrophysics, California Institute of Technology, Pasadena, CA 91125, USA}
\affiliation{Owens Valley Radio Observatory, California Institute of Technology, Big Pine, CA 93513, USA}

\author{Ivey Davis}
\affiliation{Cahill Center for Astronomy and Astrophysics, California Institute of Technology, Pasadena, CA 91125, USA}
\affiliation{Owens Valley Radio Observatory, California Institute of Technology, Big Pine, CA 93513, USA}

\author{Jayce Dowell}
\affiliation{University of New Mexico, Albuquerque, NM 87131, USA}

\author{Katherine Elder}
\affiliation{School of Earth and Space Exploration, Arizona State University, Tempe, AZ 85287, USA}

\author{Dale Gary}
\affiliation{Center for Solar-Terrestrial Research, New Jersey Institute of Technology, Newark, NJ 07102, USA}

\author{Gregg Hallinan}
\affiliation{Cahill Center for Astronomy and Astrophysics, California Institute of Technology, Pasadena, CA 91125, USA}
\affiliation{Owens Valley Radio Observatory, California Institute of Technology, Big Pine, CA 93513, USA}

\author{Charlie Harnach}
\affiliation{Owens Valley Radio Observatory, California Institute of Technology, Big Pine, CA 93513, USA}

\author{Greg Hellbourg}
\affiliation{Cahill Center for Astronomy and Astrophysics, California Institute of Technology, Pasadena, CA 91125, USA}
\affiliation{Owens Valley Radio Observatory, California Institute of Technology, Big Pine, CA 93513, USA}

\author{Jack Hickish}
\affiliation{Real-Time Radio Systems Ltd, Bournemouth, Dorset BH6 3LU, UK}

\author{Rick Hobbs}
\affiliation{Owens Valley Radio Observatory, California Institute of Technology, Big Pine, CA 93513, USA}

\author{David Hodge}
\affiliation{Cahill Center for Astronomy and Astrophysics, California Institute of Technology, Pasadena, CA 91125, USA}

\author{Mark Hodges}
\affiliation{Owens Valley Radio Observatory, California Institute of Technology, Big Pine, CA 93513, USA}

\author{Yuping Huang}
\affiliation{Cahill Center for Astronomy and Astrophysics, California Institute of Technology, Pasadena, CA 91125, USA}
\affiliation{Owens Valley Radio Observatory, California Institute of Technology, Big Pine, CA 93513, USA}

\author{Andrea Isella}
\affiliation{Department of Physics and Astronomy, Rice University, Houston, TX 77005, USA}

\author{Daniel C. Jacobs}
\affiliation{School of Earth and Space Exploration, Arizona State University, Tempe, AZ 85287, USA}

\author{Ghislain Kemby}
\affiliation{Owens Valley Radio Observatory, California Institute of Technology, Big Pine, CA 93513, USA}

\author{John T. Klinefelter}
\affiliation{Owens Valley Radio Observatory, California Institute of Technology, Big Pine, CA 93513, USA}

\author{Matthew Kolopanis}
\affiliation{School of Earth and Space Exploration, Arizona State University, Tempe, AZ 85287, USA}

\author{Nikita Kosogorov}
\affiliation{Cahill Center for Astronomy and Astrophysics, California Institute of Technology, Pasadena, CA 91125, USA}
\affiliation{Owens Valley Radio Observatory, California Institute of Technology, Big Pine, CA 93513, USA}

\author{James Lamb}
\affiliation{Owens Valley Radio Observatory, California Institute of Technology, Big Pine, CA 93513, USA}

\author{Casey Law}
\affiliation{Cahill Center for Astronomy and Astrophysics, California Institute of Technology, Pasadena, CA 91125, USA}
\affiliation{Owens Valley Radio Observatory, California Institute of Technology, Big Pine, CA 93513, USA}

\author{Nivedita Mahesh}
\affiliation{Cahill Center for Astronomy and Astrophysics, California Institute of Technology, Pasadena, CA 91125, USA}
\affiliation{Owens Valley Radio Observatory, California Institute of Technology, Big Pine, CA 93513, USA}

\author{Surajit Mondal}
\affiliation{Center for Solar-Terrestrial Research, New Jersey Institute of Technology, Newark, NJ 07102, USA}

\author{Brian O'Donnell}
\affiliation{Center for Solar-Terrestrial Research, New Jersey Institute of Technology, Newark, NJ 07102, USA}

\author{Kathryn A. Plant}
\affiliation{Owens Valley Radio Observatory, California Institute of Technology, Big Pine, CA 93513, USA}
\affiliation{Jet Propulsion Laboratory, California Institute of Technology, Pasadena, CA 91011, USA}

\author{Corey Posner}
\affiliation{Owens Valley Radio Observatory, California Institute of Technology, Big Pine, CA 93513, USA}

\author{Travis Powell}
\affiliation{Owens Valley Radio Observatory, California Institute of Technology, Big Pine, CA 93513, USA}

\author{Vinand Prayag}
\affiliation{Owens Valley Radio Observatory, California Institute of Technology, Big Pine, CA 93513, USA}

\author{Andres Rizo}
\affiliation{Owens Valley Radio Observatory, California Institute of Technology, Big Pine, CA 93513, USA}

\author{Andrew Romero-Wolf}
\affiliation{Jet Propulsion Laboratory, California Institute of Technology, Pasadena, CA 91011, USA}

\author{Jun Shi}
\affiliation{Cahill Center for Astronomy and Astrophysics, California Institute of Technology, Pasadena, CA 91125, USA}

\author{Greg Taylor}
\affiliation{University of New Mexico, Albuquerque, NM 87131, USA}

\author{Jordan Trim}
\affiliation{Owens Valley Radio Observatory, California Institute of Technology, Big Pine, CA 93513, USA}

\author{Mike Virgin}
\affiliation{Owens Valley Radio Observatory, California Institute of Technology, Big Pine, CA 93513, USA}

\author{Akshatha Vydula}
\affiliation{School of Earth and Space Exploration, Arizona State University, Tempe, AZ 85287, USA}

\author{Sandy Weinreb}
\affiliation{Cahill Center for Astronomy and Astrophysics, California Institute of Technology, Pasadena, CA 91125, USA}

\author{Scott White}
\affiliation{Owens Valley Radio Observatory, California Institute of Technology, Big Pine, CA 93513, USA}

\author{David Woody}
\affiliation{Owens Valley Radio Observatory, California Institute of Technology, Big Pine, CA 93513, USA}

\author{Sijie Yu}
\affiliation{Center for Solar-Terrestrial Research, New Jersey Institute of Technology, Newark, NJ 07102, USA}

\author{Thomas Zentmeyer}
\affiliation{Owens Valley Radio Observatory, California Institute of Technology, Big Pine, CA 93513, USA}


\begin{abstract}
Strong solar activities are often accompanied by a variety of radio bursts. These radio bursts not only serve as valuable diagnostics of coronal and heliospheric processes but also as potential tools in space weather monitoring and forecasting. However, space weather applications call for the capability for low-latency and high-sensitivity radio burst recording and reporting, which has remained lacking.
In this work, we present the development of a near-realtime radio burst recording and reporting system with the Owens Valley Radio Observatory's Long Wavelength Array. 
The system directly clips data from the realtime buffer and streams it as a live realtime radio dynamic spectrogram. The spectrograms are then fed to a deep-learning–based burst identification module for type III radio bursts. The identifier is built on a YOLO (You Only Look Once) architecture, trained by synthetic type III radio bursts generated by using a physic-based model to achieve accurate and robust detection. This system enables continuous real-time radio spectrum streaming and the automatic reporting of type III radio bursts within $\sim$10 seconds of their occurrence.

\end{abstract}


\section{Introduction} \label{sec:intro}

Radio observations are an important tool for space weather monitoring and forecasting. Solar radio bursts are among the earliest electromagnetic signatures of energetic particle acceleration and shock formation in the solar corona and heliosphere. Because many radio bursts are closely associated with solar flares, coronal mass ejections (CMEs), and the release of solar energetic particles (SEPs), they carry strong potential in space-weather monitoring and forecasting \citep{bastian2001coronal, gopalswamy2009cme, reid2018solar, white2024solar, Balch2008SpWea}. In particular, rapidly drifting type~III bursts trace energetic electron beams escaping along open magnetic field lines, while type~II bursts reveal CME-driven shocks that are primary accelerators of large SEP events \citep{kouloumvakos2019connecting, morosan2025determining}. These strong physical connections make solar radio observations particularly valuable for both scientific diagnostics and operational space-weather applications. To fully exploit this potential, however, radio burst detection must be performed automatically with very low latency and high accuracy, enabling timely identification of eruptive activity without reliance on manual inspection.

Over the past decade, major advances in low-frequency radio astronomy have been driven by new-generation interferometric arrays such as the Low Frequency Array (LOFAR; \citealt{vanHaarlem2013,Carley2020JSWSC}) and the Murchison Widefield Array (MWA; \citealt{Tingay2013}). These instruments provide high sensitivity, broad bandwidth, and high resolution in both the time--frequency and spatial domains, enabling detailed studies of solar radio bursts across a wide range of intensities and morphologies \citep[e.g.,][]{Morosan2014, Morosan2015, kouloumvakos2019connecting}. Results from these facilities have demonstrated strong connections between radio burst properties and solar eruptive phenomena, including CME kinematics, shock evolution, and coronal density structure \citep{morosan2025resolving}. Despite their scientific power, however, these arrays are primarily research-oriented. Their data products are typically processed offline, with substantial latency between observation and analysis, and they do not provide continuous real-time data streams optimized for rapid event reporting. As a result, their exceptional data quality has not yet translated into operational real-time space weather capability.

Real-time monitoring of solar radio emission has long been recognized as a critical capability for operational space-weather nowcasting and forecasting. Several established systems, such as the Radio Solar Telescope Network (RSTN) \citep{NOAA_NCEI_solar_radio_datasets,RSTN_1981}, e-Callisto \citep{ecallisto_2009} can provide full-day coverage of dynamic spectra that have enabled decades of research on burst occurrence and flare–CME associations. 
However, these systems typically exhibit latency at the minute scale between realtime acquisition and operation/public availability, and their sensitivity and spectral resolution are often insufficient to capture the faint, rapidly evolving radio bursts now known to accompany many weak events or as precursors of major events \citep{Klein2018CRPhy}. 
Currently, despite the wealth of broadband instrumentation currently available, there remains a substantial gap between data acquisition at the telescope and low-latency, high-sensitivity burst reporting suitable for immediate scientific response or operational space-weather use.

The Owens Valley Radio Observatory's Long Wavelength Array (OVRO-LWA) \citep{OVROLWA} occupies a unique position between these two regimes. 
\revpz{OVRO-LWA has 352 cross-dipole antennas, observing over a frequency range of 13.4--86.9 MHz (instantaneous bandwidth 73.5 MHz). In its beamformer (total-power) mode, the raw frequency resolution is 24 kHz across 3072 channels, and the Level-1 spectrogram cadence is 256 ms (fast visibility at 0.1 s). In addition, time resolution in the beamformer chain can reach 1 ms (typically 64 ms), while slow visibility products provide 10 s cadence for longer-timescale monitoring.}
As a wide-field, low-frequency interferometric array, OVRO-LWA delivers high sensitivity and broadband coverage, while its beamformed total-power data stream enables continuous monitoring of solar radio emission. Crucially, the OVRO-LWA recording architecture makes it possible to access data with very low latency, opening a pathway to near-real-time spectral and imaging monitoring. When combined with reliable automatic detection, such a system has the potential to detect and report solar radio bursts within seconds of their onset, providing a powerful new capability for both scientific response and operational space-weather nowcasting and forecasting.

With the development of large low-frequency array infrastructures, the data volume and rate of production of observation data has grown to the point where manual inspection is infeasible.
There are early attempts of using thresholding algorithms to identify solar radio bursts \citep{lobzin2009automatic,zhang2018type}.  Machine-learning methods---especially deep object-detection networks---have recently shown superior performance in identifying diverse solar radio burst morphologies \citep{scully2023improved,he2023solar,deng2024real}. 
Yet a major remaining challenge is the development of a complete end-to-end system capable not only of accurate detection but also of robust \emph{real-time} ingestion, processing, and reporting.

In this paper, we introduce a real-time solar radio burst reporting system developed for OVRO-LWA. The system streams spectroscopic data directly from the instrument’s live buffer to a burst identification module based on YOLO (You Only Look Once \citealt{redmon2016you}) trained using a physics-based type III radio burst generation model. This integrated pipeline achieves high sensitivity while maintaining low latency, enabling automatic identification and reporting of type III radio bursts within 10 seconds of onset. Section 2 presents the real-time streaming of the dynamic spectrum, and Section 3 presents the event detection method. Section 4 is the conclusion and discussion.

\section{Realtime Spectrum System Framework}

\begin{figure}[h!]
    \centering
    \includegraphics[width=0.99\linewidth]{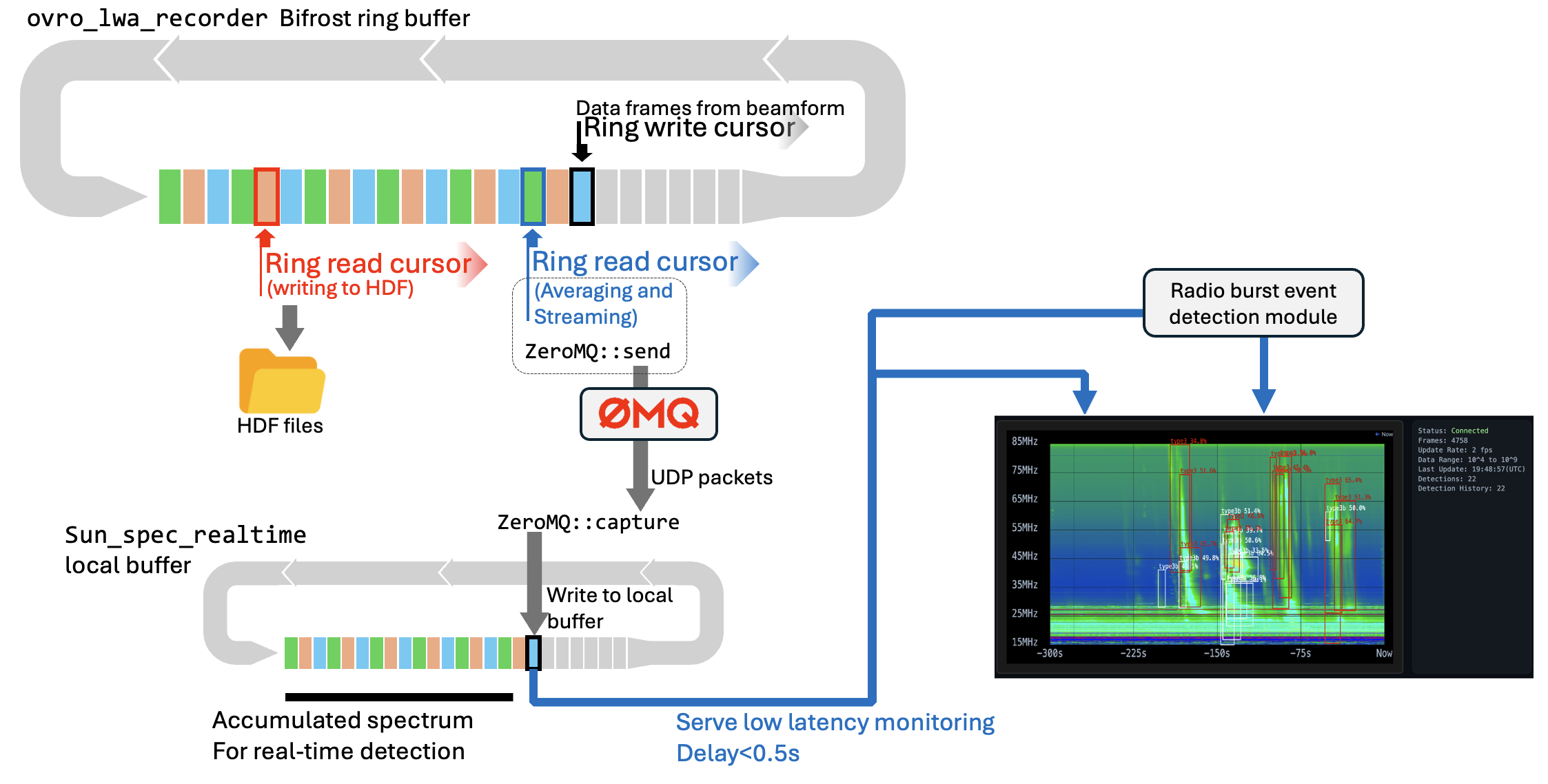}
    \caption{    Conceptual workflow of the OVRO-LWA real-time spectrum system.
    The \texttt{ovro\_lwa\_recorder} Bifrost ring buffer receives
    beamformed data frames and is read out along two independent
    cursors: one writes HDF5 files for the standard archive, while
    the other performs averaging and streaming via ZeroMQ.
    On the subscriber side, \texttt{Sun\_spec\_realtime} accumulates
    spectra in a local ring buffer and serves low-latency monitoring
    to the event–detection module.}
    \label{fig:concept}
\end{figure}

The real-time spectrum system is designed to sit non-invasively on top
of the existing OVRO--LWA recording infrastructure while providing a
low-latency, data stream for monitoring and automatic burst
detection.  Figure~\ref{fig:concept} summarizes the overall workflow.
At the front end, the \texttt{ovro\_lwa\_recorder} process receives
beamformed total-power data and push into a Bifrost ring buffer \citep{cranmer2017bifrost}.

Two independent read cursors then operate on this shared buffer.
The first cursor corresponds to the production data path and writes
time-ordered blocks to on-disk HDF5 files for long-term archiving and
offline analysis.  This path is unchanged from the standard OVRO--LWA
solar recording pipeline.
The second read cursor provides the real-time branch of the system with ZeroMQ messaging framework\footnote{ZeroMQ \url{https://zeromq.org/}}.
It continuously reads short blocks from the Bifrost ring buffer,
performs light averaging and formatting, and publishes the resulting
spectroscopic frames over a ZeroMQ Publish (\texttt{PUB}) socket.  Each transmitted frame
carries a self-describing header (time tags, array shape, and data
type) together with a binary payload containing the averaged spectrum.
On the local network, the end-to-end overhead of this averaging and
streaming stage is typically $\lesssim 0.5$~s from the time the last
sample enters the recorder ring to its appearance on the wire, which
sets the lower bound on the achievable reporting latency. 

The following is a detailed description of the components of the framework.

\textit{Data flow Topology:} The producer publishes multi-part ZeroMQ messages over a \texttt{PUB} socket; consumers connect via subscribe \texttt{SUB} sockets using topic-based filtering (e.g., \texttt{``spec/solar''}). On the producer, a bounded ring buffer (shared memory or Temporary File System (tmpfs)-backed) decouples I/O from network transmission; high-water marks (HWM) on both endpoints bound memory and enforce backpressure. To preserve real-time behavior, frames older than the most recent window are dropped (drop-tail) once the HWM is reached, favoring latency over completeness.

\textit{Message composition:} Each transmission is a three-part message:
\begin{enumerate}\setlength\itemsep{0.25em}
  \item {Topic} (ASCII): stream name (e.g., \texttt{spec/ovro-lwa}).
  \item {Header} (UTF-8 JSON): a self-describing metadata dictionary (Table~\ref{tab:zmq_header}).
  \item {Payload} (binary): a contiguous little-endian \texttt{float32} array of shape \(\mathrm{data\_shape}=(N_{\rm beam},N_{\rm pol},N_{\rm chan},N_t)\).
\end{enumerate}
Headers are JSON-encoded for human readability and broad client compatibility; the payload remains binary for throughput. Optional compression (e.g., \texttt{zstd}) can be enabled when bandwidth is constrained, at the expense of a few milliseconds of CPU time.

\textit{Header schema and timing:}
Table~\ref{tab:zmq_header} lists the required fields. In particular, \texttt{time\_tag} is the instrument-native tag for the first sample in the payload; \texttt{last\_block\_time} provides a human-readable UTC derived from the acquisition clock (converted with \texttt{LWATime}); \texttt{timestamp} records the wall-clock time of message creation on the producer. Together, these fields allow end-to-end latency auditing (from \texttt{last\_block\_time} to downstream alert emission) and unambiguous indexing across independent consumers.

\begin{table}[ht]
\centering
\caption{Spectrum Streaming Data Packet Header}
\label{tab:zmq_header}
\begin{tabular}{lll}
\hline
\hline
\textbf{Field} & \textbf{Type} & \textbf{Description} \\
\hline
\texttt{time\_tag} & string/int & Original data time tag from instrument header (\texttt{ihdr}). \\
\texttt{nbeam} & int & Number of beams in the data stream. \\
\texttt{nchan} & int & Number of frequency channels. \\
\texttt{npol} & int & Number of polarizations. \\
\texttt{timestamp} & string & Time of message creation (system wall clock). \\
\texttt{last\_block\_time} & string & Human-readable UTC timestamp of the last processed data block (converted from \texttt{LWATime}). \\
\texttt{data\_shape} & tuple & Shape of the averaged data array (\texttt{avg\_data.shape}). \\
\texttt{data\_type} & string & Data type encoding of the array (here, little-endian 32-bit float: \texttt{<f4}). \\
\hline
\end{tabular}
\end{table}

On the stream-subscriber side, the \texttt{Sun\_spec\_realtime} service
captures the ZeroMQ stream and writes incoming frames into a second
local ring buffer.  This buffer serves two purposes:  (1) It
supports an operator-facing display that renders continuously updated
dynamic spectra for human monitoring of solar activity with sub-second
lag.  (2) It accumulates a sliding window of spectra that is
periodically assembled into fixed-size dynamic-spectrum tiles
(e.g., updating every 10~s with overlapping windows) and passed to the radio
burst event–detection module.  \revpz{For low-latency operation, the update cadence is set to 10~s using overlapping windows so that short-lived burst onsets can be detected and localized within the near-real-time alert budget. Because the model input tile spans the full 640-pixel time axis (corresponding to ~320~s at 0.5~s per pixel), changing the 10~s cadence primarily trades alert latency against computational throughput and the likelihood of capturing very brief structures in successive windows; for applications that tolerate lower temporal cadence, larger intervals (e.g., 4--5 minutes) can be used.}  The detection back end, described in
Section~\ref{sec:ml_detection}, operates on these tiles to identify and
localize burst activity and to issue machine-readable alerts.  By
decoupling the telescope recorder, streaming layer, and detection
module through ring buffers and ZeroMQ messaging, the framework
achieves robust low-latency operation while preserving the integrity
and throughput of the primary OVRO--LWA data pipeline.

\paragraph{Data Streaming Latency}
With zero-copy payload handling, JSON header parsing, and optional compression disabled on LAN, the end-to-end streaming overhead is sub-second. Combined with pre-processing and model inference, the system meets the \(\sim 10\,\mathrm{s}\) target from \texttt{last\_block\_time} to alert for short-lived radio bursts such as type~III radio bursts, while maintaining continuous coverage for longer-lived activity.

\section{Event Detection with Machine Learning}\label{sec:ml_detection}

In order to enable reliable low-latency identification of solar radio bursts in the real-time spectroscopic stream, we develop a machine-learning detection module trained on a large synthetic dataset constructed from first-principles physical models. This section describes the physics-based event generator, background modeling, dataset construction, and the performance of the trained detector.

\subsection{Physics-based Event Generation}

Type~III and type~IIIb solar radio bursts are produced by beams of suprathermal electrons propagating along open magnetic field lines in the corona. As the beam travels outward, the local electron plasma frequency decreases monotonically with height, causing the emitted radiation to drift from high to low frequencies in the dynamic spectrum.  
For a given density profile in the corona $n_e(r)$ (\citealp[e.g.][]{saito77}), the plasma frequency $f_p \approx 9\sqrt{n_e}$~kHz maps the electron-beam trajectory to a characteristic time--frequency curve.  
We synthesize these curves using adjustable beam speeds, launch
times, and density-scale heights, enabling fine control of drift rates, durations, and bandwidths.

Fine structures---notably the striae of type~IIIb bursts---can be explained by density turbulence and small-scale inhomogeneities that fragment the beam--plasma emission into clumpy Langmuir-wave packets, producing quasi-periodic modulations in both intensity and emission frequency \citep{ReidKontar2021,ChenEtAl2018,Kontar2001}.
We implement these using clustered, short-lived sub-bursts with narrow instantaneous bandwidths and small random frequency offsets.  This produces synthetic morphology that closely resembles observed type~IIIb striation patterns.

\subsection{Brightness Statistics and Flux Distribution}

\begin{figure}[h]
    \centering
    \includegraphics[width=0.55\linewidth]{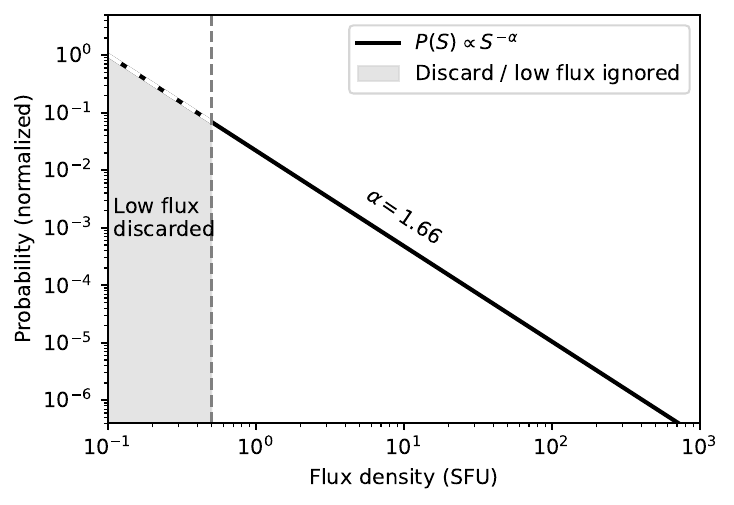}
    \caption{Flux density distribution used for synthetic burst generation.
    Weak events below the gray shaded region are not included, following a
    power-law brightness model with index $\alpha=1.66$.}
    \label{fig:flux}
\end{figure}

Solar radio burst brightness follows a long-tailed distribution, with many weak
events and comparatively few high-intensity bursts.  To reflect this, we draw burst
flux densities from a power-law distribution
$P(S) \propto S^{-\alpha}$ with $\alpha = 1.66$,  \citep{saint2012decade}
consistent with statistical studies of type~III brightness distributions.
A minimum-flux threshold is applied to remove extremely weak events that would be
indistinguishable from realistic noise fluctuations.  
Figure~\ref{fig:flux} illustrates the sampled distribution and the low-flux
cutoff used during dataset construction.

\subsection{Backgrounds, Noise, and RFI Robustness}

Real dynamic spectra from the OVRO--LWA contain substantial variability arising from
quiet-Sun spectral structure, instrumental gain fluctuations, ionospheric variations,
and intermittent radio-frequency interference (RFI).  
To ensure that the trained detector remains robust under realistic observing
conditions, we construct a background model directly from observations rather than
synthetic parametric noise.

\begin{figure}[h]
    \centering
    \includegraphics[width=0.4\linewidth]{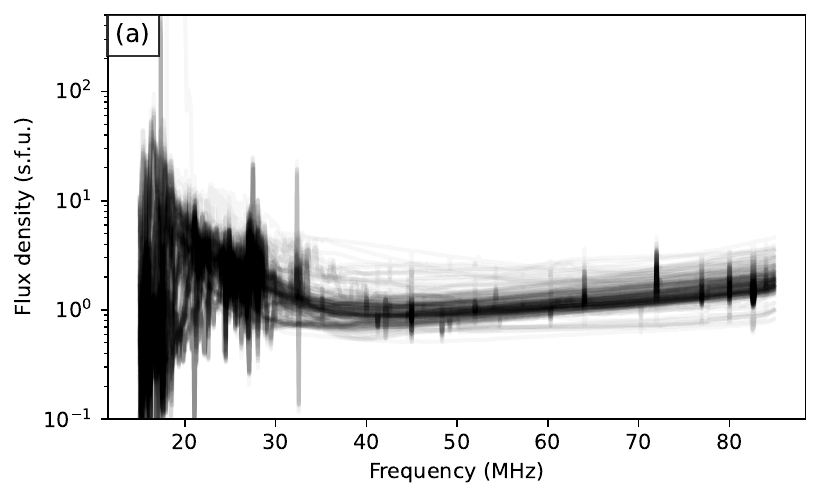}
    \includegraphics[width=0.4\linewidth]{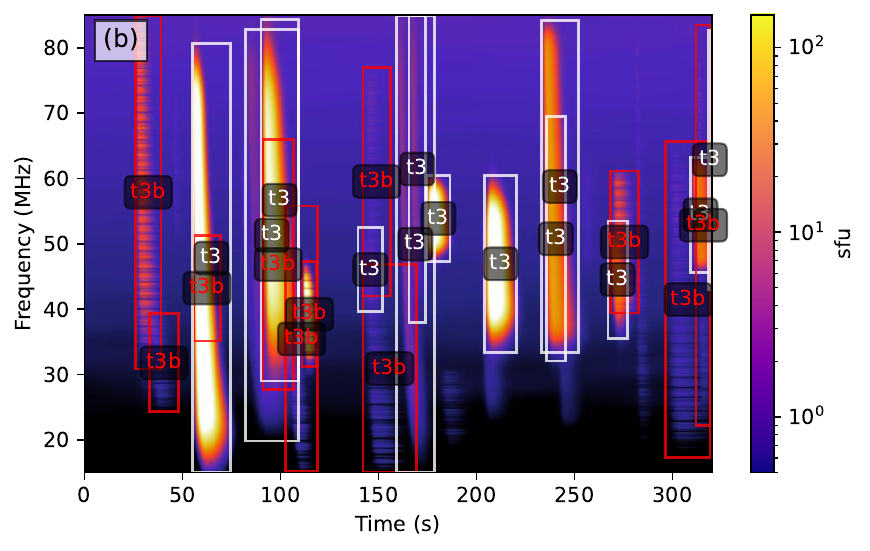}
    \includegraphics[width=0.4\linewidth]{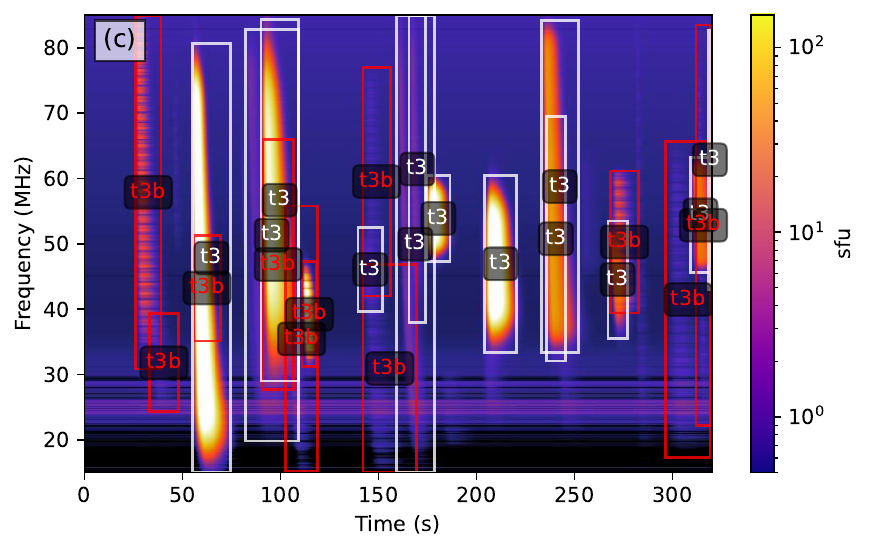}
    \includegraphics[width=0.4\linewidth]{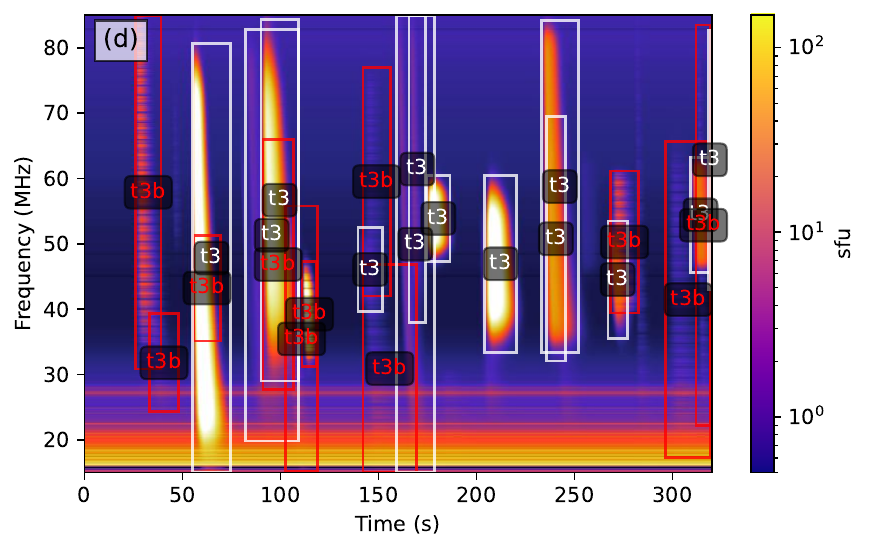}
    \caption{
    (a) Distribution of the 264 selected solar-quiet spectra used to construct the
    background library.  
    (b) Synthetic events on a modeled quiet-Sun background.  
    (c) Events injected into a relatively RFI-quiet background.  
    (d) Events injected into a strongly RFI-contaminated background.
    In panels (b) (c) (d) the type III bursts are labeled as 't3' and type IIIb bursts are labeled as 't3b'.
    }
    \label{fig:bg}
\end{figure}

We selected 264 solar-quiet full-day spectra from routine OVRO--LWA operations.
These intervals contain no strong solar burst activity, but they do
span a wide range of RFI conditions, from relatively clean to highly contaminated.
Panel~(a) of Figure~\ref{fig:bg} shows the distribution of these backgrounds
as a function of frequency: the black curves represent the measured quiet-Sun
power spectra, while the faint envelope captures the observed scatter across
different days.  This ensemble serves as the foundation for the background library.
Panels~(b)–(d) illustrate how these measured backgrounds are incorporated into the
event generator. Specifically, 
Panel~(b) shows simulated type~III and IIIb burst events superposed on a
modeled quiet-Sun brightness background, demonstrating clean contrast in the
absence of significant interference.  
Panel~(c) uses a background drawn from a relatively RFI-quiet day, producing
a realistic but moderately contaminated dynamic spectrum.  
Panel~(d) shows the same event population injected into a background from a
strongly contaminated RFI day.
By mixing events with backgrounds spanning quiet, mildly contaminated, and highly
RFI-rich conditions, the training set captures the full diversity of OVRO--LWA
observing environments.  
This substantially improves the detector’s resilience to false positives, maintains
performance in the presence of fluctuating baselines, and ensures that the
model generalizes effectively to real data encountered during live operations.

\subsection{Synthetic Dataset Construction}

Using the physics-based generator and measured background library, we produce a
dataset of 50,000 dynamic-spectrum images containing over
$1.09 \times 10^{6}$ labeled events spanning type~III and type~IIIb classes.
Each event is stored with its exact bounding box
($\{t_0, t_1, \nu_0, \nu_1\}$, which denotes the start and ending time stamps and frequencies, respectively), class label (type III or type IIIb), and associated physical parameters
(drift rate, beam speed, peak brightness).
\revpz{The full dataset is publicly released on Huggingface \footnote{Dataset DOI: \texttt{10.57967/hf/7000} \url{https://doi.org/10.57967/hf/7000}.}}
\revpz{The synthetic event generator and the training/inference pipeline used to produce the dataset are available at \texttt{PhySynthTrainer}\footnote{\url{https://github.com/peijin94/PhySynthTrainer}}.}
The dataset is split into 40,000 training, 5,000 validation, and 5,000 test
spectrograms.

\subsection{Training}

We adopt YOLOv8m~\citep{yolov8_ultralytics} for real-time inference.
Dynamic-spectrum frames are normalized using a logarithmic scale with a lower limit of 0.5 sfu (solar flux unit; 1 sfu is 10$^{-19}$~erg~s$^{-1}$~cm$^{-2}$~Hz$^{-1}$) and an upper limit of 200 sfu. They are resized to 640 by 640 pixels, and are then fed as 2D single-channel images to the YOLOv8m.
\revpz{For reference, the 640 pixels along the time axis correspond to 0.5 s per pixel, giving an effective time span of 320 s across the tile. The 640 frequency pixels cover the usable band (approximately 14--87 MHz), corresponding to roughly 0.11 MHz per frequency pixel.}
The model is trained using standard bounding-box regression, objectiveness, and classification losses with class-balancing to address the natural imbalance between type~III and type~IIIb events.

\begin{figure}[h!]
    \centering
    \includegraphics[width=0.7\textwidth, trim=5 0 5 21, clip]{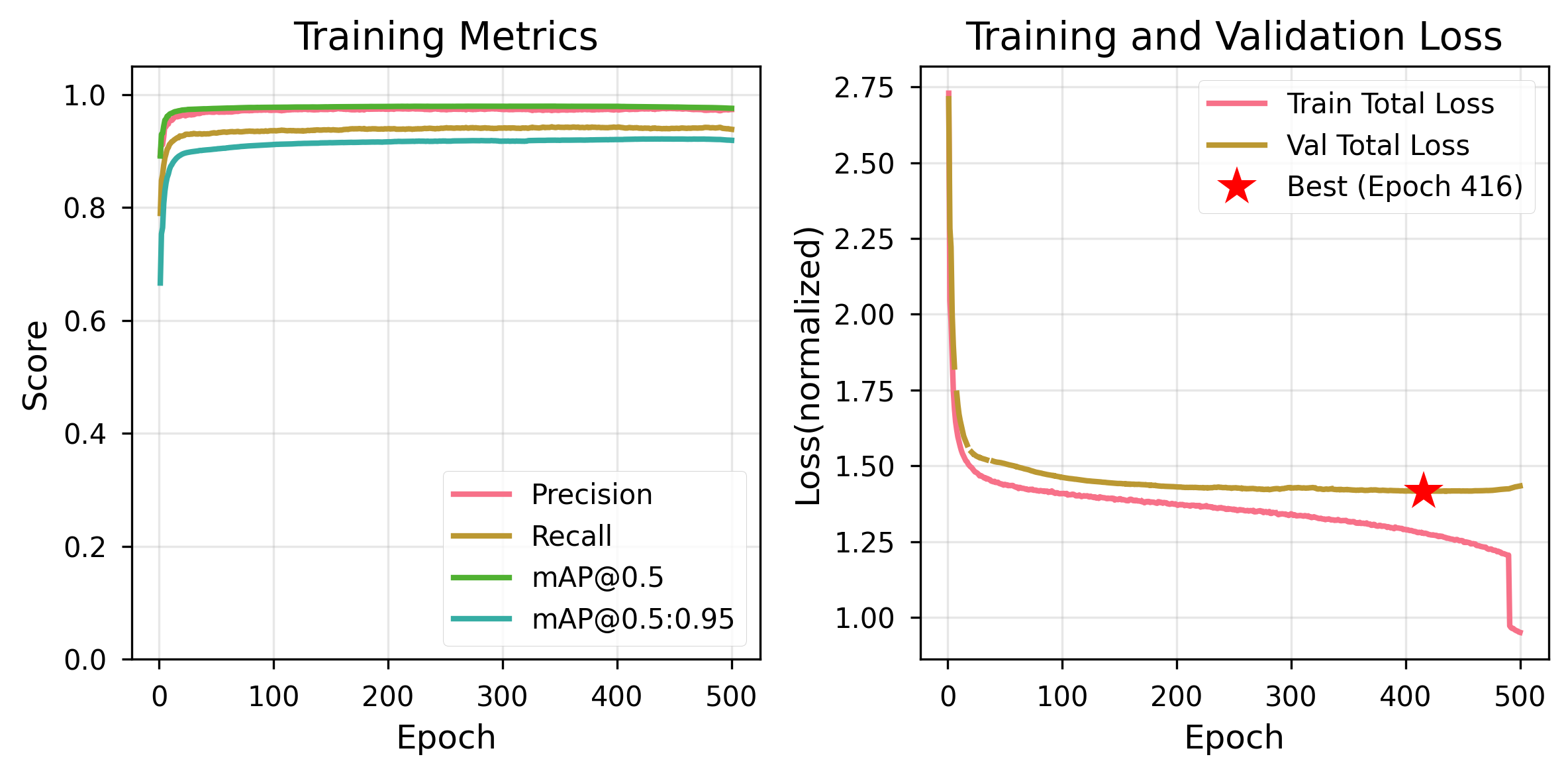}
    \caption{Training performance across four evaluation metrics (left) and total training and validation loss curves (right) using the YOLOv8m model.  The best-performing model occurs at epoch 416 (star symbol).}
    \label{fig:training}
\end{figure}

Training proceeds for up to 500 epochs, with early stopping based on validation loss.
Figure~\ref{fig:training} (left panel) shows the evolution of precision, recall,
and mean average precision (mAP) during training, all of which converge to $\gtrsim 0.9$.
(Mean average precision (mAP) is a standard object-detection metric that summarizes how well predicted bounding boxes match ground-truth labels across detection thresholds. For each class, precision--recall is computed by varying the confidence threshold, the area under that curve gives the average precision (AP), and mAP is the mean AP across classes. In this work we report mAP at a fixed IoU criterion, so higher mAP indicates more accurate localization and fewer false positives/negatives.)
The right panel shows the total training and validation losses, which decrease steadily, with the optimal
model selected at epoch~416 (marked by a star symbol).

\begin{figure}[h]
    \centering
    \includegraphics[width=0.99\linewidth, trim=0 0 0 18, clip]{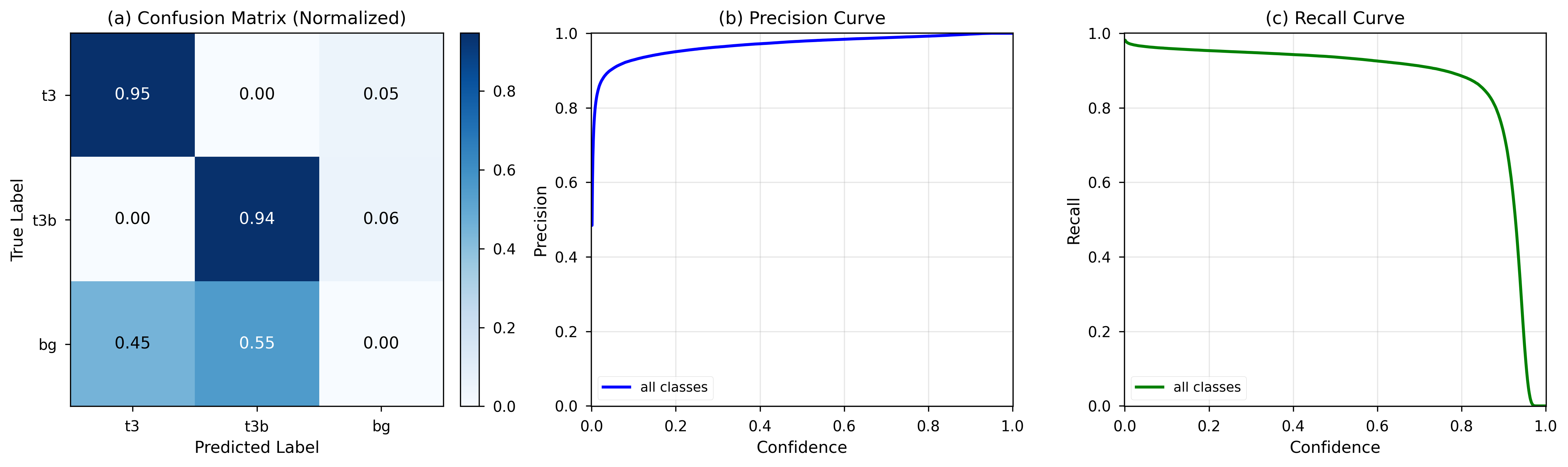}
\caption{Evaluation performance for solar radio burst Type III and Type IIIb detection. (a) Confusion matrix
    (b) Precision curve and (c) recall curve across inference confidence thresholds.}
    \label{fig:prcurve}
\end{figure}

The detector achieves high accuracy with the 5,000-image synthetic test set.
Figure~\ref{fig:prcurve} shows the confusion matrix along with precision and recall
curves.  Classification performance is strong for both burst types:
95\% of type~III bursts and 94\% of type~IIIb bursts are correctly identified.
Confusion primarily arises from faint events near the detection threshold.
\revpz{In the evaluation, background frames are not included as a separate labeled class, so any predictions on ``background'' are counted as either type~III or type~IIIb detections. This explains why the confusion matrix shows non-zero entries associated with background in the test set. From our test-set statistics, the resulting background-driven false-positive contribution corresponds to approximately 5\% for type~III and 6\% for type~IIIb. The bottom row of the confusion matrix shows how these false-positive predictions are distributed between the two burst classes, rather than indicating a 100\% false-positive rate.}
The precision–recall characteristics remain high across a wide confidence range,
confirming robust discrimination even under noisy backgrounds.

\subsection{Real-Data Evaluation and Brightness Awareness}

To evaluate the generalization of the detector with real observations, we applied the model to real OVRO-LWA spectroscopy data. Because the training set was constructed from a power-law brightness distribution, with both strong and weak bursts were generated, but only the brighter ones were labeled, the model inherits an intrinsic brightness sensitivity: it is naturally selective for moderate and strong events.

\begin{figure}[h]
    \centering
    \includegraphics[width=0.3\linewidth]{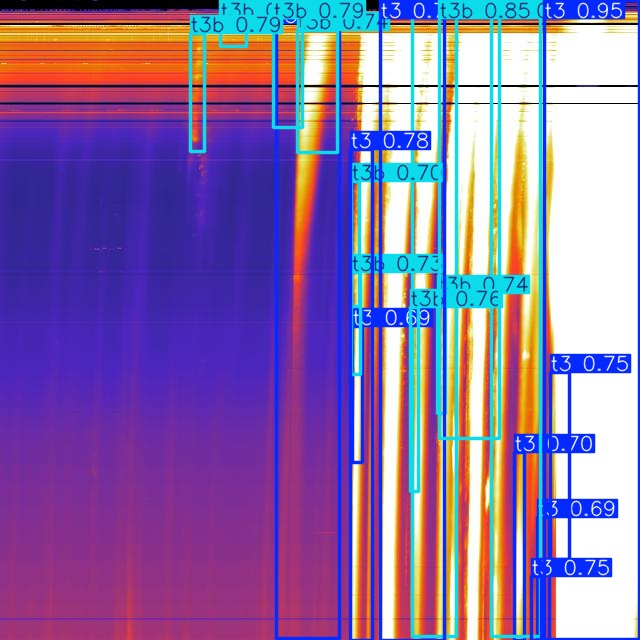}
    \includegraphics[width=0.3\linewidth]{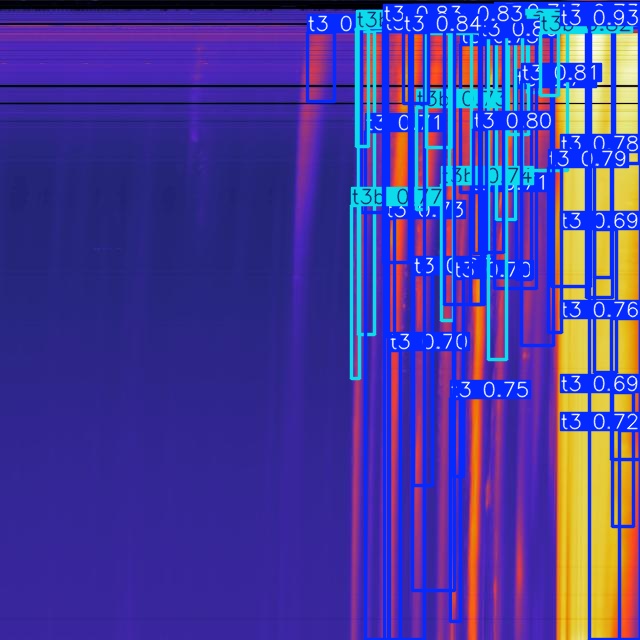}
    \includegraphics[width=0.3\linewidth]{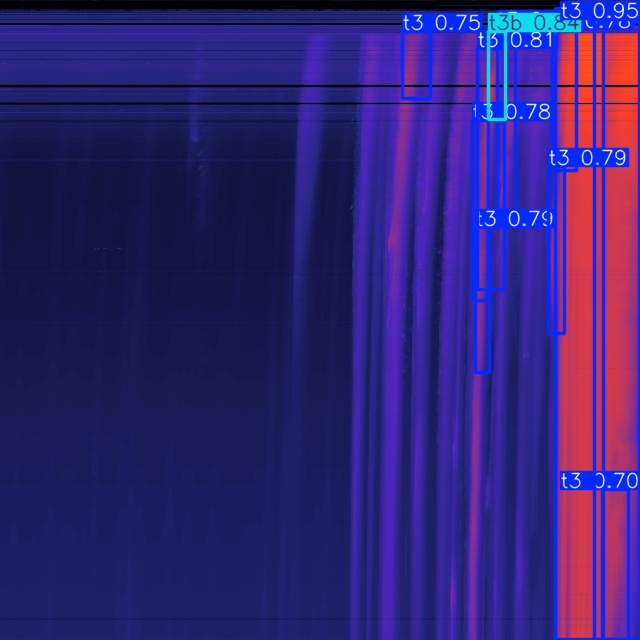}
    \caption{
    Brightness-aware behavior of the trained detector applied to real OVRO--LWA data.  From left to right: color scale ranges of 0.5--30 sfu, 0.5--200 sfu, and
    0.5--$10^{5}$ sfu.  Low-scale images highlight weak structures, mid-scale images
    match the synthetic training regime, and high-scale images emphasize only the
    brightest bursts.}
    \label{fig:brightness}
\end{figure}

\begin{figure}[h]
    \centering
    \includegraphics[width=0.5\linewidth]{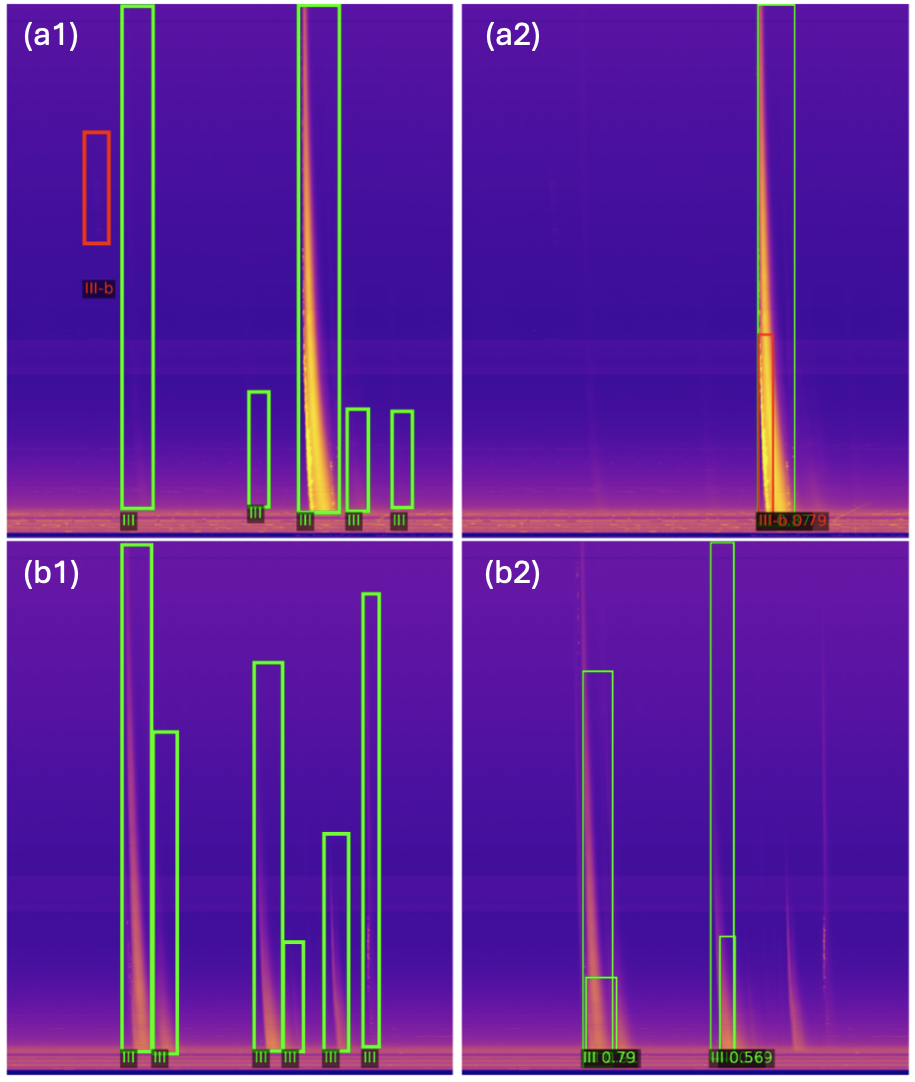}
    \caption{Example dynamic spectra with burst bounding boxes. The left column shows human annotations (panels a1, b1), and the right column shows the corresponding model detections (panels a2, b2). Green rectangles mark detected/annotated burst intervals, while the red box in (a1) highlights a type~IIIb fine-structure (striae) region. Images in this figure are scaled 0.5-150 sfu with log scale}

    \label{fig:human_model_compare}
\end{figure}

Figure~\ref{fig:brightness} illustrates this behavior using three versions of the same dynamic spectrum, displayed with different color-scale ranges:  
(1) low range (0.5--30 sfu),  
(2) medium range (0.5--200 sfu), and  
(3) high range (0.5--$10^{5}$ sfu).  
\revpz{In the operational detector, we use the medium-range scaling corresponding to the synthetic training regime; the left and right panels with 0.5--30 sfu and 0.5--$10^{5}$ sfu are shown for visualization to illustrate brightness-aware behavior across regimes.} 
These different scalings emphasize progressively brighter features in the data. 
In the low-range panel (0.5--30 sfu), faint background fluctuations and weak burst structures become visually prominent.  The model therefore produces dense detections across both weak and strong features.  The brightest burst cluster, however, is saturated and visually blended, causing the detector to label it as a single event.
\revpz{The medium-range panel (0.5--200 sfu) corresponds to the brightness regime represented in the synthetic training set. In this scaling, the weakest bursts are ignored, and the model correctly identifies the moderate and strong type~III and type~IIIb bursts with well-separated bounding boxes. In the right panel (0.5--$10^{5}$ sfu), this behavior is consistent with the configuration of the synthetic dataset; the model reports detections for only the top few brightest structures.}

\begin{figure}[]
    \centering
    \includegraphics[width=0.55\linewidth]{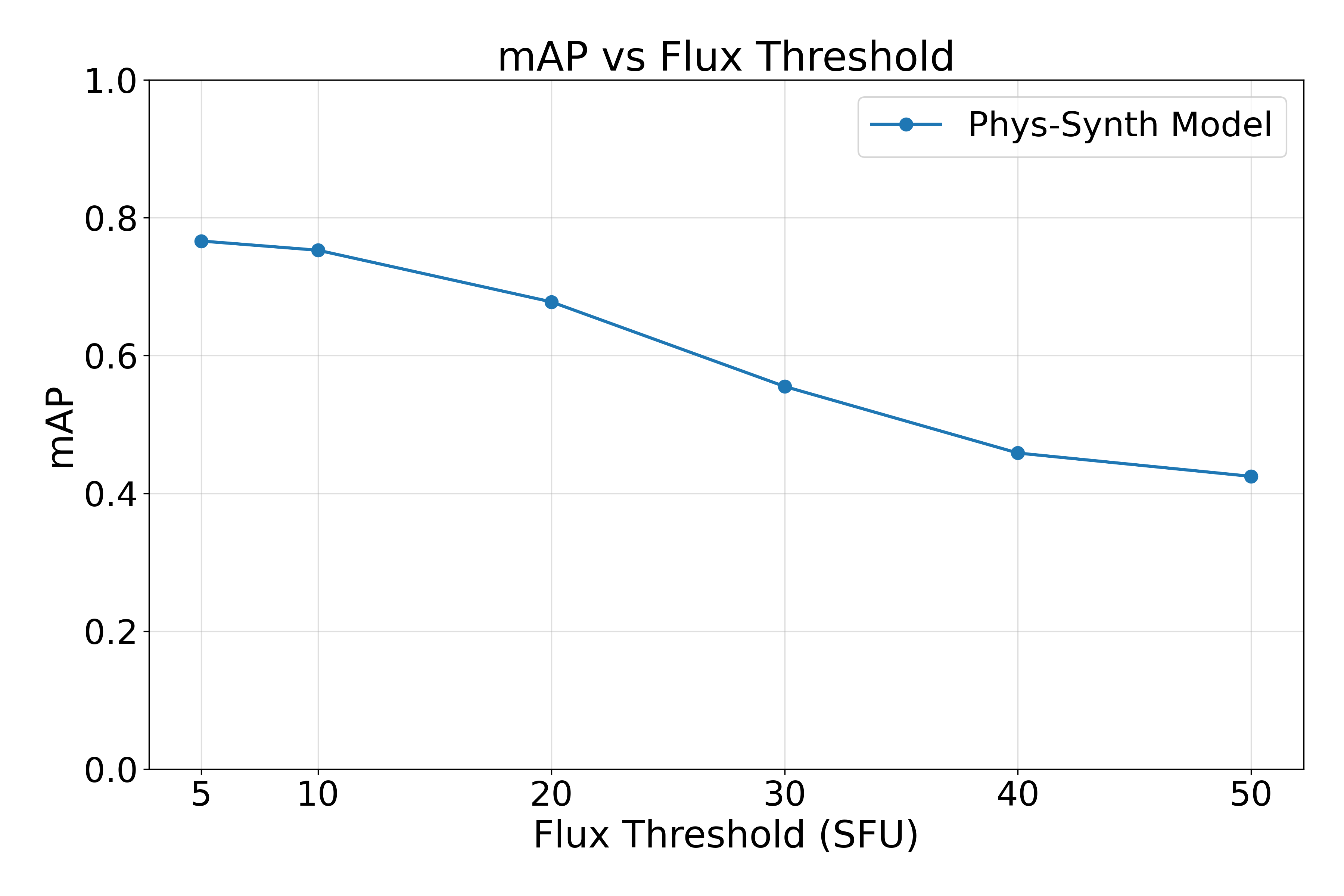}
    \caption{Detection mAP versus burst peak flux threshold. Performance decreases at high fluxes as burst morphology becomes more complex and extended, despite higher signal strength. }
    \label{fig:benchmark}
\end{figure}

We did a benchmark by applying the model on a human labeled dataset. We use Mean average precision (mAP) to evaluate the performance of the model.
{The human labeled dataset has a mAP of 0.8 at low fluxes which decreases to 0.5 mAP at the 20+ SFU flux threshold and then rises back to about 0.6 mAP above 30+ SFU.}\revpz{Figure~\ref{fig:human_model_compare} shows a qualitative comparison between the model detections and the corresponding human annotations, illustrating the model’s ability to capture both isolated events and type~IIIb fine-structure (striae) regions.}
Figure \ref{fig:benchmark} shows detection performance as a function of burst peak flux. While mAP exceeds 0.95 for faint events (~1 SFU), it decreases for the brightest bursts due to increased morphological complexity, including multi-lane structures and extended emission regions. This demonstrates that the model correctly tracks physically driven changes in burst structure rather than merely signal-to-noise.

These tests demonstrate that the detector can transfer the features in synthetic data effectively to OVRO-LWA data. Meanwhile, it behaves predictably with respect to the brightness distribution represented in training. Strong bursts can be reliably identified, weak bursts are detected when the display range emphasizes them, and saturated clusters
are merged in the same manner that a human observer would interpret the underlying spectrogram.  Overall, the model performs robustly on real data without requiring
domain adaptation.

\FloatBarrier

\subsection{Potential Scientific and Operational Applications}

\paragraph{Autonomous triggering for solar physics.}
The low-latency detection capability enables a range of automated science operations.
For instance, the system can be used to trigger rapid-response observations with instruments such as EOVSA (interrupting calibration and initiating solar observing mode) or FOXSI sounding-rocket campaigns, where even tens of seconds of advanced warning is
valuable. More broadly, the framework can facilitate coordinated multi-wavelength campaigns by providing reliable, real-time indicators of the onset of strong solar activity in the corona.

\paragraph{Space-weather monitoring and early warning.}
Type~III and type~IIIb bursts are key signatures of electron escape along open field lines and are often associated with early phases of eruptive events.  Real-time reporting of these bursts can therefore support operational space-weather services by providing near-instantaneous alerts for the onset of particle acceleration.
Compared to traditional systems such as RSTN, which provide minute-scale availability, the OVRO–LWA system offers substantially shorter delay times and higher sensitivity, making it particularly well suited for high-cadence monitoring and nowcasting applications.

\section{Conclusion and Discussion}
\label{sec:conclusion}

In this work, we have developed and demonstrated a complete end-to-end system for low-latency solar radio burst recording and reporting using OVRO-LWA solar radio observations in the meter-decameter wavelength regime of 15--87 MHz.  The system integrates real-time spectroscopic streaming, physics-informed synthetic data generation, and a fast deep-learning-based detection module into a
robust operational framework capable of issuing automatic alerts for type III and type IIIb radio bursts within $\sim$10\,s of their occurrence.  With a dual-cursor ring-buffer architecture and a ZeroMQ-based streaming layer, the pipeline achieves sub-second transport latency, ensuring that detection and alert
generation are driven by near-real-time measurements rather than archival products.
This represents a substantial advancement over existing solar radio monitoring networks, which typically operate with minute-scale latencies and lower sensitivity.

For event detection, a central component of this effort is the construction of a large physics-informed training dataset of 50,000 synthetic dynamic spectra comprising more than $10^{6}$ labeled type~III
and type~IIIb events.  These simulations encode physically realistic electron-beam kinematics, burst morphologies, brightness statistics, and measured background spectra under a wide range of RFI conditions.  As a result, the trained detector generalizes effectively to real OVRO–LWA observations without the need for domain adaptation.  The
model exhibits predictable brightness sensitivity, reliably detecting moderate and strong events while identifying weak bursts when the color scale enhances their visibility.
Real-data tests with human labeling confirm that the detector captures both isolated bursts and complex burst clusters, even in the presence of fluctuating quiet-Sun baselines or moderate RFI.

Overall, the system presented here demonstrates that low-latency, high-sensitivity, machine-learning–assisted solar radio burst reporting is now feasible with modern wide-field low-frequency arrays.  By combining real-time data streaming with a physics-informed detector, the framework provides a foundation for both automated scientific discovery and operational space-weather support, and represents a significant
step toward fully autonomous heliophysics observing systems.

\revpz{Future work will expand the detector beyond the current type~III/type~IIIb training set. In particular, type~II bursts---key signatures of CME-driven shocks---and other burst classes (e.g., type~V) are not yet represented in the training data, so the present system should be viewed as optimized for type~III-related electron-beam activity. We are building a larger multi-type labeled dataset (including type~II, additional burst classes, and noise-storm/background scenarios). Once this labeling effort is completed, we will retrain/fine-tune the detector using these well-observed events to better reflect real burst morphology and to enable a direct comparison between observed- and synthetic-training performance.}

\section{Acknowledgements}
P. Z. acknowledges support for this research by the NASA Living with a Star Jack Eddy Postdoctoral Fellowship Program, administered by UCAR’s Cooperative Programs for the Advancement of Earth System Science (CPAESS) under award 80NSSC22M0097. 
This work is supported by Google Cloud for computing resource support. This material is based upon work supported by the Google Cloud Research Credits program with the award GCP19980904.
The OVRO-LWA expansion project was supported by NSF under grant AST-1828784. OVRO-LWA operations for solar and space weather sciences are supported by NSF under grant AGS-2436999. 





%

\vspace{5mm}
\facilities{OVRO-LWA}


\software{astropy, pytorch}



\bibliography{cite}{}

@article{ReidKontar2021,
  author  = {Reid, Hamish A. S. and Kontar, Eduard P.},
  title   = {Fine structure of type III solar radio bursts from Langmuir wave motion in turbulent plasma},
  journal = {Nature Astronomy},
  volume  = {5},
  number  = {8},
  pages   = {796--804},
  year    = {2021},
  doi     = {10.1038/s41550-021-01370-8}
}

@article{ChenEtAl2018,
  author  = {Chen, Xingyao and Kontar, Eduard P. and Yu, Sijie and Yan, Yihua and Huang, Jing and Tan, Baolin},
  title   = {Fine Structures of Solar Radio Type III Bursts and Their Possible Relationship with Coronal Density Turbulence},
  journal = {The Astrophysical Journal},
  volume  = {856},
  number  = {1},
  pages   = {73},
  year    = {2018},
  doi     = {10.3847/1538-4357/aaa9bf}
}

@article{Kontar2001,
  author  = {Kontar, Eduard P.},
  title   = {Dynamics of electron beams in the solar corona plasma with density fluctuations},
  journal = {Astronomy and Astrophysics},
  volume  = {375},
  pages   = {629--637},
  year    = {2001},
  doi     = {10.1051/0004-6361:20010807}
}

@ARTICLE{Balch2008SpWea,
       author = {{Balch}, Christopher C.},
        title = "{Updated verification of the Space Weather Prediction Center's solar energetic particle prediction model}",
      journal = {Space Weather},
         year = 2008,
        month = jan,
       volume = {6},
       number = {1},
          eid = {S01001},
        pages = {S01001},
          doi = {10.1029/2007SW000337},
       adsurl = {https://ui.adsabs.harvard.edu/abs/2008SpWea...6.1001B}
}

@ARTICLE{Carley2020JSWSC,
       author = {{Carley}, Eoin P. and {Baldovin}, Carla and {Benthem}, Pieter and {Bisi}, Mario M. and {Fallows}, Richard A. and {Gallagher}, Peter T. and {Olberg}, Michael and {Rothkaehl}, Hanna and {Vermeulen}, Rene and {Vilmer}, Nicole and {Barnes}, David},
        title = "{Radio observatories and instrumentation used in space weather science and operations}",
      journal = {Journal of Space Weather and Space Climate},
         year = 2020,
        month = jan,
       volume = {10},
          eid = {7},
        pages = {7},
          doi = {10.1051/swsc/2020007},
       adsurl = {https://ui.adsabs.harvard.edu/abs/2020JSWSC..10....7C}
}

@ARTICLE{Klein2018CRPhy,
       author = {{Klein}, Karl-Ludwig and {Matamoros}, Carolina Salas and {Zucca}, Pietro},
        title = "{Solar radio bursts as a tool for space weather forecasting}",
      journal = {Comptes Rendus Physique},
         year = 2018,
        month = jan,
       volume = {19},
       number = {1-2},
        pages = {36-42},
          doi = {10.1016/j.crhy.2018.01.005},
       adsurl = {https://ui.adsabs.harvard.edu/abs/2018CRPhy..19...36K}
}

@inproceedings{redmon2016you,
  title={You only look once: Unified, real-time object detection},
  author={Redmon, Joseph and Divvala, Santosh and Girshick, Ross and Farhadi, Ali},
  booktitle={Proceedings of the IEEE conference on computer vision and pattern recognition},
  pages={779--788},
  year={2016}
}

@article{cranmer2017bifrost,
  title={Bifrost: A Python/C++ framework for high-throughput stream processing in astronomy},
  author={Cranmer, Miles D and Barsdell, Benjamin R and Price, Danny C and Dowell, Jayce and Garsden, Hugh and Dike, Veronica and Eftekhari, Tarraneh and Hegedus, Alexander M and Malins, Joseph and Obenberger, Kenneth S and others},
  journal={Journal of Astronomical Instrumentation},
  volume={6},
  number={04},
  pages={1750007},
  year={2017},
  publisher={World Scientific}
}

@ARTICLE{saito77,
       author = {{Saito}, K. and {Poland}, A.~I. and {Munro}, R.~H.},
        title = "{A study of the background corona near solar minimum.}",
      journal = {\solphys},
         year = 1977,
        month = nov,
       volume = {55},
       number = {1},
        pages = {121-134},
          doi = {10.1007/BF00150879},
       adsurl = {https://ui.adsabs.harvard.edu/abs/1977SoPh...55..121S}
}

@misc{NOAA_NCEI_solar_radio_datasets,
  author       = {{National Centers for Environmental Information (NCEI), NOAA}},
  title        = {Solar Radio Datasets},
  howpublished = {\url{https://www.ncei.noaa.gov/products/space-weather/legacy-data/solar-radio-datasets}},
  year         = 2025,
  note         = {Accessed: 2025-11-17}
}

@article{saint2012decade,
  title={A decade of solar type III radio bursts observed by the Nan{\c{c}}ay radioheliograph 1998--2008},
  author={Saint-Hilaire, Pascal and Vilmer, Nicole and Kerdraon, Alain},
  journal={The Astrophysical Journal},
  volume={762},
  number={1},
  pages={60},
  year={2012},
  publisher={IOP Publishing}
}

@INPROCEEDINGS{OVROLWA,
       author = {{Hallinan}, Gregg and {Anderson}, Marin and {Isella}, Andrea and {Gary}, Dale and {Bowman}, Judd and {Romero-Wolf}, Andrew and {OVRO-LWA Collaboration}},
        title = "{The OVRO-LWA array {\textemdash} upgraded for real-time continuous all-sky imaging from 12-85 MHz}",
    booktitle = {American Astronomical Society Meeting Abstracts \#241},
         year = 2023,
       series = {American Astronomical Society Meeting Abstracts},
       volume = {241},
        month = jan,
          eid = {451.09},
        pages = {451.09},
       adsurl = {https://ui.adsabs.harvard.edu/abs/2023AAS...24145109H}
}

@ARTICLE{ecallisto_2009,
       author = {{Benz}, A.~O. and {Monstein}, C. and {Meyer}, H. and {Manoharan}, P.~K. and {Ramesh}, R. and {Altyntsev}, A. and {Lara}, A. and {Paez}, J. and {Cho}, K.-S.},
        title = "{A World-Wide Net of Solar Radio Spectrometers: e-CALLISTO}",
      journal = {Earth Moon and Planets},
         year = 2009,
        month = apr,
       volume = {104},
       number = {1-4},
        pages = {277-285},
          doi = {10.1007/s11038-008-9267-6},
       adsurl = {https://ui.adsabs.harvard.edu/abs/2009EM&P..104..277B}
}

@INPROCEEDINGS{RSTN_1981,
       author = {{Guidice}, D.~A. and {Cliver}, E.~W. and {Barron}, W.~R. and {Kahler}, S.},
        title = "{The Air Force RSTN System}",
    booktitle = {Bulletin of the American Astronomical Society},
         year = 1981,
       volume = {13},
        month = mar,
        pages = {553},
       adsurl = {https://ui.adsabs.harvard.edu/abs/1981BAAS...13Q.553G}
}

@ARTICLE{vanHaarlem2013,
 author = {{van Haarlem}, M.~P. and {Wise}, M.~W. and {Gunst}, A.~W. and {Heald}, G. and {McKean}, J.~P. and {Hessels}, J.~W.~T. and {de Bruyn}, A.~G. and {Nijboer}, R. and {Swinbank}, J. and {Fallows}, R. and {Brentjens}, M. and {Nelles}, A. and {Beck}, R. and {Falcke}, H. and {Fender}, R. and {H{\"o}randel}, J. and {Koopmans}, L.~V.~E. and {Mann}, G. and {Miley}, G. and {R{\"o}ttgering}, H. and {Stappers}, B.~W. and {Wijers}, R.~A.~M.~J. and {Zaroubi}, S. and {van den Akker}, M. and {Alexov}, A. and {Anderson}, J. and {Anderson}, K. and {van Ardenne}, A. and {Arts}, M. and {Asgekar}, A. and {Avruch}, I.~M. and {Batejat}, F. and {B{\"a}hren}, L. and {Bell}, M.~E. and {Bell}, M.~R. and {van Bemmel}, I. and {Bennema}, P. and {Bentum}, M.~J. and {Bernardi}, G. and {Best}, P. and {B{\^\i}rzan}, L. and {Bonafede}, A. and {Boonstra}, A.-J. and {Braun}, R. and {Bregman}, J. and {Breitling}, F. and {van de Brink}, R.~H. and {Broderick}, J. and {Broekema}, P.~C. and {Brouw}, W.~N. and {Br{\"u}ggen}, M. and {Butcher}, H.~R. and {van Cappellen}, W. and {Ciardi}, B. and {Coenen}, T. and {Conway}, J. and {Coolen}, A. and {Corstanje}, A. and {Damstra}, S. and {Davies}, O. and {Deller}, A.~T. and {Dettmar}, R.-J. and {van Diepen}, G. and {Dijkstra}, K. and {Donker}, P. and {Doorduin}, A. and {Dromer}, J. and {Drost}, M. and {van Duin}, A. and {Eisl{\"o}ffel}, J. and {van Enst}, J. and {Ferrari}, C. and {Frieswijk}, W. and {Gankema}, H. and {Garrett}, M.~A. and {de Gasperin}, F. and {Gerbers}, M. and {de Geus}, E. and {Grie{\ss}meier}, J.-M. and {Grit}, T. and {Gruppen}, P. and {Hamaker}, J.~P. and {Hassall}, T. and {Hoeft}, M. and {Holties}, H.~A. and {Horneffer}, A. and {van der Horst}, A. and {van Houwelingen}, A. and {Huijgen}, A. and {Iacobelli}, M. and {Intema}, H. and {Jackson}, N. and {Jelic}, V. and {de Jong}, A. and {Juette}, E. and {Kant}, D. and {Karastergiou}, A. and {Koers}, A. and {Kollen}, H. and {Kondratiev}, V.~I. and {Kooistra}, E. and {Koopman}, Y. and {Koster}, A. and {Kuniyoshi}, M. and {Kramer}, M. and {Kuper}, G. and {Lambropoulos}, P. and {Law}, C. and {van Leeuwen}, J. and {Lemaitre}, J. and {Loose}, M. and {Maat}, P. and {Macario}, G. and {Markoff}, S. and {Masters}, J. and {McFadden}, R.~A. and {McKay-Bukowski}, D. and {Meijering}, H. and {Meulman}, H. and {Mevius}, M. and {Middelberg}, E. and {Millenaar}, R. and {Miller-Jones}, J.~C.~A. and {Mohan}, R.~N. and {Mol}, J.~D. and {Morawietz}, J. and {Morganti}, R. and {Mulcahy}, D.~D. and {Mulder}, E. and {Munk}, H. and {Nieuwenhuis}, L. and {van Nieuwpoort}, R. and {Noordam}, J.~E. and {Norden}, M. and {Noutsos}, A. and {Offringa}, A.~R. and {Olofsson}, H. and {Omar}, A. and {Orr{\'u}}, E. and {Overeem}, R. and {Paas}, H. and {Pandey-Pommier}, M. and {Pandey}, V.~N. and {Pizzo}, R. and {Polatidis}, A. and {Rafferty}, D. and {Rawlings}, S. and {Reich}, W. and {de Reijer}, J.-P. and {Reitsma}, J. and {Renting}, G.~A. and {Riemers}, P. and {Rol}, E. and {Romein}, J.~W. and {Roosjen}, J. and {Ruiter}, M. and {Scaife}, A. and {van der Schaaf}, K. and {Scheers}, B. and {Schellart}, P. and {Schoenmakers}, A. and {Schoonderbeek}, G. and {Serylak}, M. and {Shulevski}, A. and {Sluman}, J. and {Smirnov}, O. and {Sobey}, C. and {Spreeuw}, H. and {Steinmetz}, M. and {Sterks}, C.~G.~M. and {Stiepel}, H.-J. and {Stuurwold}, K. and {Tagger}, M. and {Tang}, Y. and {Tasse}, C. and {Thomas}, I. and {Thoudam}, S. and {Toribio}, M.~C. and {van der Tol}, B. and {Usov}, O. and {van Veelen}, M. and {van der Veen}, A.-J. and {ter Veen}, S. and {Verbiest}, J.~P.~W. and {Vermeulen}, R. and {Vermaas}, N. and {Vocks}, C. and {Vogt}, C. and {de Vos}, M. and {van der Wal}, E. and {van Weeren}, R. and {Weggemans}, H. and {Weltevrede}, P. and {White}, S. and {Wijnholds}, S.~J. and {Wilhelmsson}, T. and {Wucknitz}, O. and {Yatawatta}, S. and {Zarka}, P. and {Zensus}, A.},
        title = "{LOFAR: The LOw-Frequency ARray}",
      journal = {\aap},
         year = 2013,
        month = aug,
       volume = {556},
          eid = {A2},
        pages = {A2},
          doi = {10.1051/0004-6361/201220873},
archivePrefix = {arXiv},
       eprint = {1305.3550},
 primaryClass = {astro-ph.IM},
       adsurl = {https://ui.adsabs.harvard.edu/abs/2013A&A...556A...2V}
}

@ARTICLE{Tingay2013,
 author = {{Tingay}, S.~J. and {Goeke}, R. and {Bowman}, J.~D. and {Emrich}, D. and {Ord}, S.~M. and {Mitchell}, D.~A. and {Morales}, M.~F. and {Booler}, T. and {Crosse}, B. and {Wayth}, R.~B. and {Lonsdale}, C.~J. and {Tremblay}, S. and {Pallot}, D. and {Colegate}, T. and {Wicenec}, A. and {Kudryavtseva}, N. and {Arcus}, W. and {Barnes}, D. and {Bernardi}, G. and {Briggs}, F. and {Burns}, S. and {Bunton}, J.~D. and {Cappallo}, R.~J. and {Corey}, B.~E. and {Deshpande}, A. and {Desouza}, L. and {Gaensler}, B.~M. and {Greenhill}, L.~J. and {Hall}, P.~J. and {Hazelton}, B.~J. and {Herne}, D. and {Hewitt}, J.~N. and {Johnston-Hollitt}, M. and {Kaplan}, D.~L. and {Kasper}, J.~C. and {Kincaid}, B.~B. and {Koenig}, R. and {Kratzenberg}, E. and {Lynch}, M.~J. and {Mckinley}, B. and {Mcwhirter}, S.~R. and {Morgan}, E. and {Oberoi}, D. and {Pathikulangara}, J. and {Prabu}, T. and {Remillard}, R.~A. and {Rogers}, A.~E.~E. and {Roshi}, A. and {Salah}, J.~E. and {Sault}, R.~J. and {Udaya-Shankar}, N. and {Schlagenhaufer}, F. and {Srivani}, K.~S. and {Stevens}, J. and {Subrahmanyan}, R. and {Waterson}, M. and {Webster}, R.~L. and {Whitney}, A.~R. and {Williams}, A. and {Williams}, C.~L. and {Wyithe}, J.~S.~B.},
        title = "{The Murchison Widefield Array: The Square Kilometre Array Precursor at Low Radio Frequencies}",
      journal = {\pasa},
         year = 2013,
        month = jan,
       volume = {30},
          eid = {e007},
        pages = {e007},
          doi = {10.1017/pasa.2012.007},
archivePrefix = {arXiv},
       eprint = {1206.6945},
 primaryClass = {astro-ph.IM},
       adsurl = {https://ui.adsabs.harvard.edu/abs/2013PASA...30....7T}
}

@ARTICLE{Morosan2014,
author = {{Morosan}, D.~E. and {Gallagher}, P.~T. and {Zucca}, P. and {Fallows}, R. and {Carley}, E.~P. and {Mann}, G. and {Bisi}, M.~M. and {Kerdraon}, A. and {Konovalenko}, A.~A. and {MacKinnon}, A.~L. and {Rucker}, H.~O. and {Thid{\'e}}, B. and {Magdaleni{\'c}}, J. and {Vocks}, C. and {Reid}, H. and {Anderson}, J. and {Asgekar}, A. and {Avruch}, I.~M. and {Bentum}, M.~J. and {Bernardi}, G. and {Best}, P. and {Bonafede}, A. and {Bregman}, J. and {Breitling}, F. and {Broderick}, J. and {Br{\"u}ggen}, M. and {Butcher}, H.~R. and {Ciardi}, B. and {Conway}, J.~E. and {de Gasperin}, F. and {de Geus}, E. and {Deller}, A. and {Duscha}, S. and {Eisl{\"o}ffel}, J. and {Engels}, D. and {Falcke}, H. and {Ferrari}, C. and {Frieswijk}, W. and {Garrett}, M.~A. and {Grie{\ss}meier}, J. and {Gunst}, A.~W. and {Hassall}, T.~E. and {Hessels}, J.~W.~T. and {Hoeft}, M. and {H{\"o}randel}, J. and {Horneffer}, A. and {Iacobelli}, M. and {Juette}, E. and {Karastergiou}, A. and {Kondratiev}, V.~I. and {Kramer}, M. and {Kuniyoshi}, M. and {Kuper}, G. and {Maat}, P. and {Markoff}, S. and {McKean}, J.~P. and {Mulcahy}, D.~D. and {Munk}, H. and {Nelles}, A. and {Norden}, M.~J. and {Orru}, E. and {Paas}, H. and {Pandey-Pommier}, M. and {Pandey}, V.~N. and {Pietka}, G. and {Pizzo}, R. and {Polatidis}, A.~G. and {Reich}, W. and {R{\"o}ttgering}, H. and {Scaife}, A.~M.~M. and {Schwarz}, D. and {Serylak}, M. and {Smirnov}, O. and {Stappers}, B.~W. and {Stewart}, A. and {Tagger}, M. and {Tang}, Y. and {Tasse}, C. and {Thoudam}, S. and {Toribio}, C. and {Vermeulen}, R. and {van Weeren}, R.~J. and {Wucknitz}, O. and {Yatawatta}, S. and {Zarka}, P.},
        title = "{LOFAR tied-array imaging of Type III solar radio bursts}",
      journal = {\aap},
         year = 2014,
        month = aug,
       volume = {568},
          eid = {A67},
        pages = {A67},
          doi = {10.1051/0004-6361/201423936},
archivePrefix = {arXiv},
       eprint = {1407.4385},
 primaryClass = {astro-ph.SR},
       adsurl = {https://ui.adsabs.harvard.edu/abs/2014A&A...568A..67M}
}

@ARTICLE{Morosan2015,
 author = {{Morosan}, D.~E. and {Gallagher}, P.~T. and {Zucca}, P. and {O'Flannagain}, A. and {Fallows}, R. and {Reid}, H. and {Magdaleni{\'c}}, J. and {Mann}, G. and {Bisi}, M.~M. and {Kerdraon}, A. and {Konovalenko}, A.~A. and {MacKinnon}, A.~L. and {Rucker}, H.~O. and {Thid{\'e}}, B. and {Vocks}, C. and {Alexov}, A. and {Anderson}, J. and {Asgekar}, A. and {Avruch}, I.~M. and {Bentum}, M.~J. and {Bernardi}, G. and {Bonafede}, A. and {Breitling}, F. and {Broderick}, J.~W. and {Brouw}, W.~N. and {Butcher}, H.~R. and {Ciardi}, B. and {de Geus}, E. and {Eisl{\"o}ffel}, J. and {Falcke}, H. and {Frieswijk}, W. and {Garrett}, M.~A. and {Grie{\ss}meier}, J. and {Gunst}, A.~W. and {Hessels}, J.~W.~T. and {Hoeft}, M. and {Karastergiou}, A. and {Kondratiev}, V.~I. and {Kuper}, G. and {van Leeuwen}, J. and {McKay-Bukowski}, D. and {McKean}, J.~P. and {Munk}, H. and {Orru}, E. and {Paas}, H. and {Pizzo}, R. and {Polatidis}, A.~G. and {Scaife}, A.~M.~M. and {Sluman}, J. and {Tasse}, C. and {Toribio}, M.~C. and {Vermeulen}, R. and {Zarka}, P.},
        title = "{LOFAR tied-array imaging and spectroscopy of solar S bursts}",
      journal = {\aap},
         year = 2015,
        month = aug,
       volume = {580},
          eid = {A65},
        pages = {A65},
          doi = {10.1051/0004-6361/201526064},
archivePrefix = {arXiv},
       eprint = {1507.07496},
 primaryClass = {astro-ph.SR},
       adsurl = {https://ui.adsabs.harvard.edu/abs/2015A&A...580A..65M}
}

@article{reid2018solar,
 author = {{Reid}, Hamish A.~S. and {Kontar}, Eduard P.},
        title = "{Solar type III radio burst time characteristics at LOFAR frequencies and the implications for electron beam transport}",
      journal = {\aap},
         year = 2018,
        month = jun,
       volume = {614},
          eid = {A69},
        pages = {A69},
          doi = {10.1051/0004-6361/201732298},
archivePrefix = {arXiv},
       eprint = {1802.01507},
 primaryClass = {astro-ph.SR},
       adsurl = {https://ui.adsabs.harvard.edu/abs/2018A&A...614A..69R}
}

@article{deng2024real,
  title={Real-time automated detection of multi-category solar radio bursts},
  author={Deng, Jingyu and Yuan, Guowu and Zhou, Hao and Wu, Hao and Tan, Chengming},
  journal={Astrophysics and Space Science},
  volume={369},
  number={10},
  pages={99},
  year={2024},
  publisher={Springer}
}

@article{scully2023improved,
  title={Improved Type III solar radio burst detection using congruent deep learning models},
  author={Scully, Jeremiah and Flynn, Ronan and Gallagher, Peter T and Carley, Eoghan P and Daly, Mark},
  journal={Astronomy \& Astrophysics},
  volume={674},
  pages={A218},
  year={2023},
  publisher={EDP Sciences}
}

@article{he2023solar,
   author = {{He}, Hailan and {Yuan}, Guowu and {Zhou}, Hao and {Tan}, Chengming and {Guo}, Shaojie},
        title = "{Solar Radio Burst Detection Based on the MobileViT-SSDLite Lightweight Model}",
      journal = {\apjs},
         year = 2023,
        month = dec,
       volume = {269},
       number = {2},
          eid = {51},
        pages = {51},
          doi = {10.3847/1538-4365/ad036c},
       adsurl = {https://ui.adsabs.harvard.edu/abs/2023ApJS..269...51H}
}

@article{zhang2018type,
  title={A type III radio burst automatic analysis system and statistic results for a half solar cycle with Nan{\c{c}}ay Decameter Array data},
  author={Zhang, PJ and Wang, Chuan Bing and Ye, Lin},
  journal={Astronomy \& Astrophysics},
  volume={618},
  pages={A165},
  year={2018},
  publisher={EDP sciences}
}

@article{morosan2025resolving,
  title={Resolving spatial and temporal shock structures using LOFAR observations of type II radio bursts},
  author={Morosan, DE and Jebaraj, IC and Zhang, P and Zucca, P and Dabrowski, B and Gallagher, PT and Krankowski, A and Vocks, C and Vainio, R},
  journal={Astronomy \& Astrophysics},
  volume={695},
  pages={A70},
  year={2025},
  publisher={EDP Sciences}
}

@software{yolov8_ultralytics,
  author = {Glenn Jocher and Ayush Chaurasia and Jing Qiu},
  title = {Ultralytics YOLOv8},
  version = {8.0.0},
  year = {2023},
  orcid = {0000-0001-5950-6979, 0000-0002-7603-6750, 0000-0003-3783-7069},
  license = {AGPL-3.0}
}

@article{bastian2001coronal,
  title={The coronal mass ejection of 1998 April 20: direct imaging at radio wavelengths},
  author={Bastian, TS and Pick, M and Kerdraon, A and Maia, D and Vourlidas, A},
  journal={The Astrophysical Journal},
  volume={558},
  number={1},
  pages={L65},
  year={2001},
  publisher={IOP Publishing}
}

@article{morosan2025determining,
  title={Determining the acceleration regions of in situ electrons using remote radio and X-ray observations},
  author={Morosan, DE and Dresing, N and Palmroos, C and Gieseler, J and Jebaraj, IC and Warmuth, A and Fedeli, A and Normo, S and Pomoell, J and Kilpua, EKJ and others},
  journal={Astronomy \& Astrophysics},
  volume={693},
  pages={A296},
  year={2025},
  publisher={EDP sciences}
}

@article{lobzin2009automatic,
  title={Automatic recognition of type III solar radio bursts: automated radio burst identification system method and first observations},
  author={Lobzin, Vasili V and Cairns, Iver H and Robinson, Peter A and Steward, Graham and Patterson, Garth},
  journal={Space Weather},
  volume={7},
  number={4},
  year={2009},
  publisher={Wiley Online Library}
}

@article{gopalswamy2009cme,
  title={CME interactions with coronal holes and their interplanetary consequences},
  author={Gopalswamy, N and M{\"a}kel{\"a}, P and Xie, H and Akiyama, S and Yashiro, S},
  journal={Journal of Geophysical Research: Space Physics},
  volume={114},
  number={A3},
  year={2009},
  publisher={Wiley Online Library}
}

@article{kouloumvakos2019connecting,
  title={Connecting the properties of coronal shock waves with those of solar energetic particles},
  author={Kouloumvakos, Athanasios and Rouillard, Alexis P and Wu, Yihong and Vainio, Rami and Vourlidas, Angelos and Plotnikov, Illya and Afanasiev, Alexandr and {\"O}nel, Hakan},
  journal={The Astrophysical Journal},
  volume={876},
  number={1},
  pages={80},
  year={2019},
  publisher={IOP Publishing}
}

@article{white2024solar,
  title={Solar radio bursts and space weather},
  author={White, Stephen M},
  journal={arXiv preprint arXiv:2405.00959},
  year={2024}
}
\bibliographystyle{aasjournal}



\end{document}